\documentclass[aps,pre,reprint,superscriptaddress]{revtex4-2}
\pdfoutput=1

\usepackage{graphicx}
\usepackage{amsmath}
\usepackage{siunitx}
\usepackage{verbatim}
\usepackage{amsfonts}
\usepackage{physics}
\usepackage{todo}

\begin{document}

%Title of paper
\title{Methods for Extremely Sparse-Angle Proton Tomography}

\author{B.~T.~Spiers}
 \email{benjamin.spiers@physics.ox.ac.uk}
\author{R.~Aboushelbaya}
\author{Q.~Feng}
\author{M.~W.~Mayr}
\author{I.~Ouatu}
\author{R.~W.~Paddock}
\author{R.~Timmis}
\author{R.~H.~W.~Wang}
\affiliation{Department of Physics, Atomic and Laser Physics sub-Department, University of Oxford, Clarendon Laboratory, Parks Road, Oxford OX1 3PU, United Kingdom}
\author{P.~A.~Norreys}
\affiliation{Department of Physics, Atomic and Laser Physics sub-Department, University of Oxford, Clarendon Laboratory, Parks Road, Oxford OX1 3PU, United Kingdom}
\affiliation{Central Laser Facility, UKRI-STFC Rutherford Appleton Laboratory, Harwell Campus, Didcot, Oxfordshire OX11 0QX, United Kingdom} 
\affiliation{John Adams Institute, Denys Wilkinson Building, Oxford OX1 3RH, United Kingdom}

\date{\today}

\newcommand{\etal}[2]{#1 \emph{et al}\cite{#2}}

\sisetup{separate-uncertainty=true,multi-part-units=single}

\begin{abstract}
Proton radiography is a widely-fielded diagnostic used to measure magnetic structures in plasma. The deflection of protons with multi-MeV kinetic energy by the magnetic fields is used to infer their path-integrated field strength. Here, the use of tomographic methods is proposed for the first time to lift the degeneracy inherent in these path-integrated measurements, allowing full reconstruction of spatially resolved magnetic field structures in three dimensions. Two techniques are proposed which improve the performance of tomographic reconstruction algorithms in cases with severely limited numbers of available probe beams, as is the case in laser-plasma interaction experiments where the probes are created by short, high-power laser pulse irradiation of secondary foil targets. The methods are equally applicable to optical probes such as shadowgraphy and interferometry [M. Kasim et al. Phys. Rev. E \textbf{95}, 023306 (2017)], thereby providing a disruptive new approach to three dimensional imaging across the physical sciences and engineering disciplines.
\end{abstract}

\maketitle

\section{Introduction}

Proton radiography is a widely-fielded diagnostic used to measure magnetic structures in plasma~\cite{borghesi,archie,chikang,chikang2,volpe,rygg,palmer,tubman,mackinnon,kugland_2013,morita,park,huntington,tzeferacos,mackinnon_2006,sarri}. A laser-generated proton beam~\cite{mackinnon_2001} is directed through a plasma, and is deflected as it propagates by local electromagnetic field structures. After exiting the plasma, the beam propagates over a distance before impinging on a detector screen. CR-39 and radiochromic film are commonly used as proton detectors and each has its own advantages and disadvantages~\cite{scullion}.

While the resulting images are difficult to relate directly to the fields in the plama~\cite{kugland}, theoretical work from \textcite{kasim1} and \textcite{archie} present algorithms which are able to recover transverse magnetic field components, path-integrated along the directions of proton probing. More recently, \textcite{kasim2} derived a statistical approach to compensate for lack of information regarding the transverse profile of the proton beam prior to interaction with the plasma. \textcite{chen} investigated the application of machine learning methods to the problem of proton radiography inversion, noting the degeneracy involved in interpreting path-integrated measurements, and suggested taking proton radiographs from multiple view angles as a method for resolving field structures spatially. While some experiments---for example those of \textcite{chikang} and more recently \textcite{tubman}---have probed similar interactions along different axes, the first full exploration of the possibility of recovering spatially resolved magnetic field structures from proton radiographs using standard tomography techniques is presented here.

In Section~\ref{sec:PR} a brief summary of proton radiography is presented, along with the theory of inverting proton radiographs and recent advances in radiographic inversion techniques. The reader is then introduced to the subject of tomography in Section~\ref{sec:tomography}, along with the \emph{filtered back-projection} algorithm (FBP), which is one of the most important and widely-used in tomography applications. In Section~\ref{sec:hankel} an analytic approach to tomography using Fourier decomposition in the angular variable is derived and a method of implementing it using filtered back-projection is presented for the first time. When implemented in this way the new approach is realised by an interpolation in observation angle. Section~\ref{sec:aspect} presents another new method, which improves reconstruction quality of functions with much larger extent in one dimension than the others by making them appear `squashed' into a more uniform aspect ratio before the FBP algorithm is used. In Section~\ref{sec:applications} it is shown how these modifications improve the quality of reconstruction for a function representing the magnetic field of a plasma channel. Section~\ref{sec:conclusion} summarizes the results, illustrates areas for further research, and concludes the article.

\section{Proton Radiography}
\label{sec:PR}

Proton radiography, in the limit of paraxiality and small in-plasma deflections, measures line-integrated magnetic fields, and is sensitive to field components transverse to the direction of probing (this is easily seen from the Lorentz force, which does not depend on the magnetic field component parallel to particle velocity). Protons launched from a distance \(z_s\) behind the target plasma are deflected by electromagnetic fields in the plasma and then travel a further distance \(z_i\) before being recorded on a detector screen. Momentum deflections \(\Delta{\vec{p}}\) experienced by protons launched in the \(z\)-direction, under the often-employed assumption that deflections are caused solely by a magnetic field \(\vec{B}\) and employing the paraxial and thin-target approximations within the plasma, are given by
\begin{align}
	\Delta{\vec{p}}\qty(\vec{x_p}) &= e\int{\hat{z} \cross \vec{B}\qty(x_p,y_p,z) \dd{z}}\\
	&= e \grad_\perp{\int{A_z\qty(x_p,y_p,z) \dd{z}}} \nonumber
\end{align}
where \(\vec{x_p} = [x_p, y_p]\) represent transverse coordinates at the plasma and \(e\) represents the fundamental unit of charge. The result is simplified by introducing the magnetic vector potential defined by \(\vec{B} = \curl{\vec{A}}\)

As a result of these deflections the protons reach positions \(\vec{X}\) on the detection screen given by
\begin{align}
	\vec{X}\qty(\vec{x_p}) &= \frac{z_s + z_i}{z_s} \vec{x_p} + \frac{\Delta{\vec{p}}}{p_0} z_i \\
	        &= M \qty( \vec{x_p} + f_g \frac{\Delta{\vec{p}}}{p_0} ). \nonumber
\end{align}

Introduced here are the magnification \(M = 1 + {z_i}/{z_s}\) and the ``geometric focal length" parameter \(f_g = \qty(z_s^{-1} + z_i^{-1})^{-1} = z_i / M\) which conveniently separate the effects of source divergence from those of deflections caused by the plasma. Image-plane structures depend on \(f_g\) as a geometric parameter, with \(M\) only affecting the overall scale of the image. The effect of the position deflections is to increase the fluence of protons in some regions and reduce it in others. The image-plane proton fluence is given by
\begin{equation}
	\Psi\qty(\vec{X}) = \sum_{\vec{X}\qty(\vec{x_p}) = \vec{X}} \Psi_0\qty(\vec{x_p}) \det(J)^{-1},
\end{equation}
where \(J\) is the Jacobian matrix for the transformation from plasma- to image-plane coordinates:
\begin{equation}
	J\qty(\vec{x_p}) = \pdv{(X,Y)}{(x_p, y_p)},
\end{equation}
whose determinant is given by
\begin{align}
	\det(J) = M^2 &\left[\qty(1 + \frac{f_g}{p_0} \pdv{\Delta{p_x}}{x_p}) \qty(1 + \frac{f_g}{p_0} \pdv{\Delta{p_y}}{y_p})\right. \\
                     &- \left.\left(\frac{f_g}{p_0}\right)^2 \pdv{\Delta{p_x}}{y_p}\pdv{\Delta{p_y}}{x_p}\right] \nonumber
\end{align}

This result can be expressed in terms of the longitudinal components of the MHD current \(\vec{j}\) and the Hessian determinant of the vector potential \(\vec{A}\):
\begin{align}
    \det(J)/M^2 &= 1 - \frac{f_g e}{\mu_0} \int{j_z\dd{z}}\label{eq:detJ}\\
                &+ f_g^2e^2 \det(H\qty(\int{A_z \dd{z}})).\nonumber
\end{align}

As argued by \textcite{archie}, in the limit of small deflections Equation~\ref{eq:detJ} may be taken to first order in \(f_g\) and used to recover the integrated longitudinal MHD current. In regimes of stronger deflection---quantified by \textcite{archie} using a contrast parameter equivalent to \(\mu = \frac{f_g \Delta{\vec{p}}}{\ell p_0}\) for plasmas with typical transverse spatial scale of variation \(\ell\)---this is not feasible and the resulting images are not a simple mathematical function of the measured fields, though iterative numerical algorithms based on solving the Monge-Amp\`{e}re problem are available~\cite{problem} which may be used to recover these fields from a proton fluence distribution.

Even an ideal reconstruction can only produce information about line-integrated fields. Symmetry assumptions may be used to make conclusions about the three-dimensional distribution of fields (for example, using Abel transform inversion), and \textcite{chen} proposed that the additional information available when taking proton radiographs from multiple different probe directions could enable full reconstruction of three-dimensional fields. This has the form of a \emph{transverse vector tomography problem}.

\section{Tomography}
\label{sec:tomography}

\subsection{Basic Theory of Tomography}

In this work we consider \emph{parallel-probe} tomography (i.e. tomography in which each observation is made using a collimated beam). While this is not strictly true due to the nature of proton beams and the processes by which they are produced in laser-plasma experiments (target normal sheath acceleration and capsule implosion both produce divergent proton beams with small virtual source located close to the point from which protons are accelerated), the distance between proton source and plasma is usually significantly larger than the transverse size of the plasma. The variation in incident proton angle over the transverse extent of the plasma is therefore sufficiently small that they may be treated as collimated for the purposes of tomographic reconstruction.

A three-dimensional scalar function \(f\qty(x,y,z)\) is defined in Cartesian coordinates. A tomographic projection of this function is parametrised by the probe angle \(\theta\). For a given \(\theta\) we define a new Cartesian coordinate system \((q, s, t)\) related to \((x,y,z)\) by rotation about the \(z\) axis: 

\begin{align}
\begin{bmatrix}
q \\
s \\
t
\end{bmatrix} &=
\begin{bmatrix}
x \cos\theta + y \sin\theta \\
y \cos\theta - x \sin\theta \\
z
\end{bmatrix}\\
\begin{bmatrix}
x \\
y \\
z
\end{bmatrix} &=
\begin{bmatrix}
q \cos\theta - s \sin\theta \\
s \cos\theta + q \sin\theta \\
t
\end{bmatrix}.
\end{align}

Taking projections along the local \(q\) direction produces for each \(\theta\) a function of \(s\) and \(t\):

\begin{align}
F_{\theta}\qty(s, t) &= \int_{-\infty}^{\infty} f\qty( q \cos{\theta} - s \sin{\theta}, s\cos{\theta} + q \sin{\theta} , t) \dd{q} \nonumber\\
   &= \mathcal{R}_{\theta} \left[f(x, y, z)\right](s, t) \nonumber\\
   &= \mathcal{R}_{\theta} \left[f(x, y, t)\right](s). \label{eq:radon}
\end{align}

This equation defines the Radon transform \(\mathcal{R}_{\theta}\), the integral transform that is the theoretical basis of tomographic analysis. It is important to note that, due to our assumption of parallel probing, this is effectively a `stack' of two-dimensional tomographs, one for each value in \(t\). The values of \(F_\theta\qty(s, t)\) at fixed \(t\) are only influenced by the two-dimensional slice of the original function for which \(z = t\): \(f\qty(x, y, t)\), for all \(s\) and \(\theta\). This allows for application of two-dimensional inversion algorithms to the three-dimensional tomography problem.

\begin{figure*}
    \includegraphics[width=\textwidth]{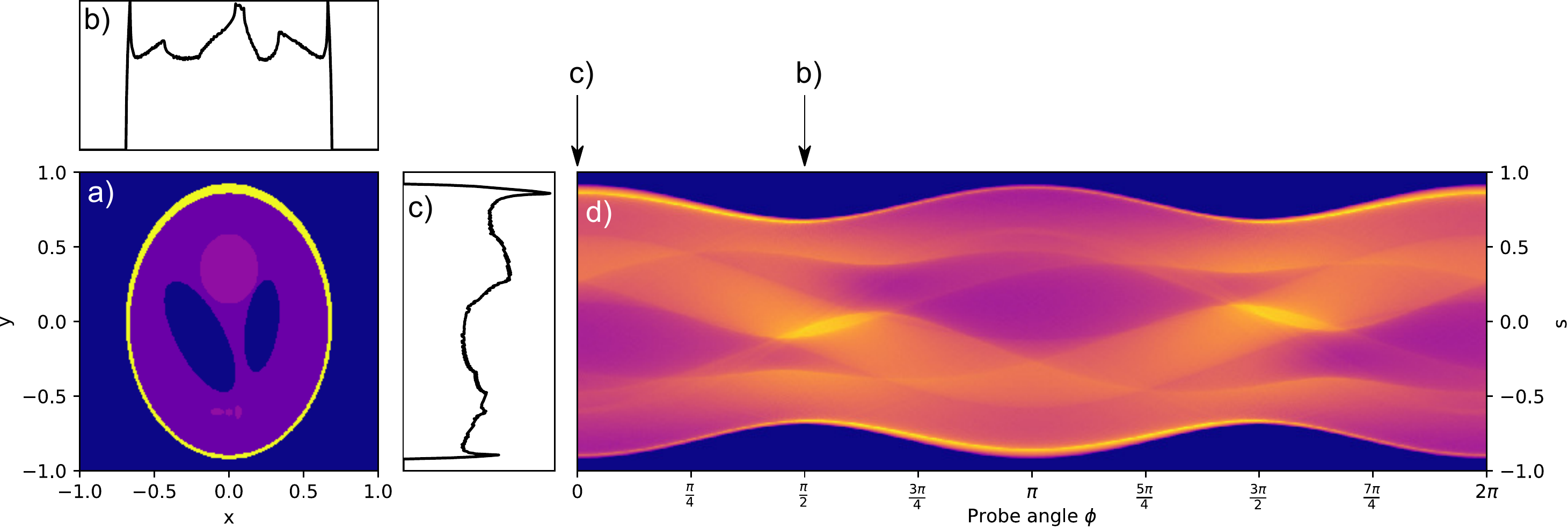}
    \caption{Pane a): The Shepp-Logan phantom \(f\qty(x,y)\) and its sinogram \(F\qty(\phi, s)\) (pane d). Panes b) and c) are projections of \(f\) along the \(y\) and \(x\) axes respectively, corresponding to \(F\qty(\frac{\pi}{2}, s)\) and \(F\qty(0, s)\). Note that \(F\) is \(2\pi\)-periodic and has the parity property \(F\qty(\phi+\pi, s) = F\qty(\phi, -s)\).}
    \label{fig:sinogram}
\end{figure*}

The function \(F_{\theta}\qty(s)\) (whose dependence on t has been dropped following the previous argument) is often visualised as a `sinogram': the parameter \(\theta\) is promoted to a variable and the resulting two-variable function \(F\qty(s, \theta)\) is plotted as an image. An example of a sinogram is shown in Figure~\ref{fig:sinogram}. This sinogram was computed from the Shepp-Logan phantom, a function often used to test tomographic techniques~\cite{shepplogan}. A modified version of the Shepp-Logan phantom with both positive and negative values is employed in the following sections, as this better represents the nature of magnetic field structures.

\subsection{The Filtered Back-Projection Algorithm}

\begin{figure}
	\includegraphics[width=\columnwidth]{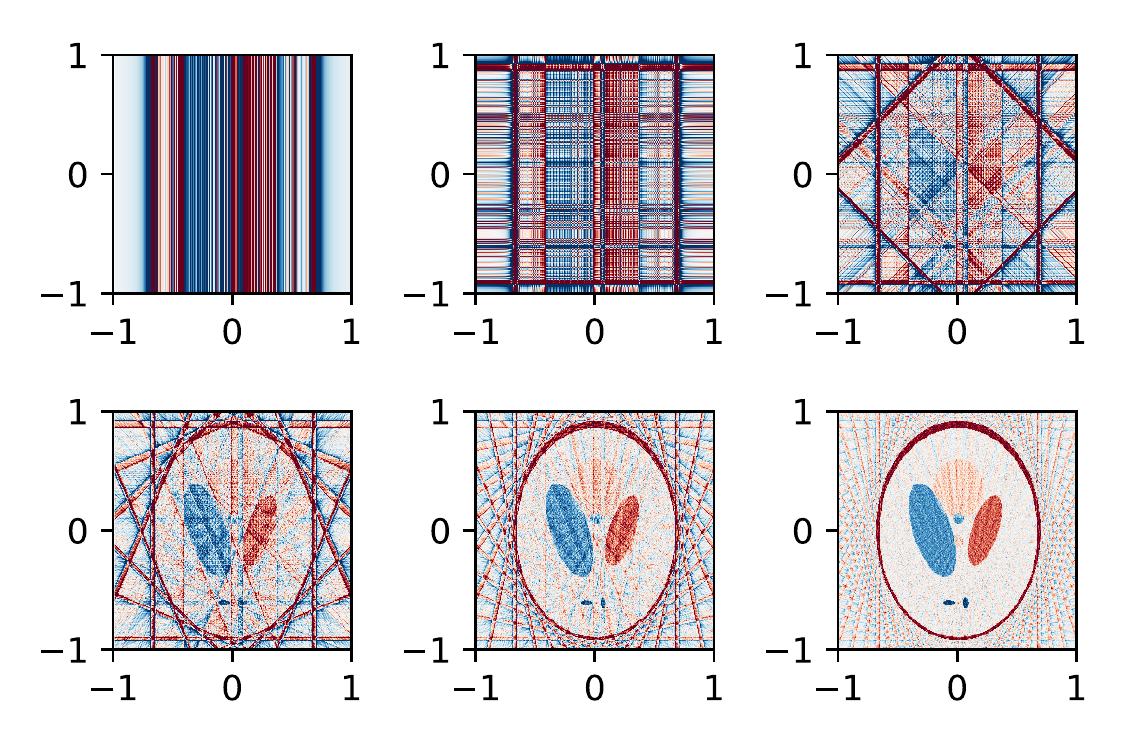}
	\caption{Example of the failure mode of FBP. 1, 2, 4, 8, 16 and 32 projections are used to produce these images. Linear streaking dominates the reconstruction for \(N \leq 4\). Even for \(N > 8\), while the original function is discernable, the exterior of the image continues to show a chequerboard-like pattern.}
	\label{fig:fbp_streaks}
\end{figure}

A canonical algorithm for the recovery of tomographic data sets is the \emph{filtered back-projection} (FBP) method. In short, this method filters each projection with a kernel proportional to \(\abs{k}\) in Fourier space, then `smears' the filtered projections across their probe directions and sums the resulting `back-projected' functions to recover an approximation of the original function. FBP converges to an analytically correct result in the limit of many projections (and can be derived as a discretisation of the Fourier projection-slice theorem discussed in Section~\ref{sec:hankel:fourier}), but where samples are few and sparse in the angular dimension it suffers from severe `streaking' artefacts. This behaviour is demonstrated in Figure~\ref{fig:fbp_streaks}.

\subsection{Tomography of Vector Functions}
In this section, consider a vector function \(\vec{g}\qty(x,y,z)\). Tomographic projections of this function are taken using the \((q, s, t)\), rather than the \((x, y, z)\), components---i.e. the components measured rotate with the angle of probing rather than being fixed in the background coordinate system.

\emph{Longitudinal vector tomography}, realised for example by Doppler tomography of fluid velocity fields, measures the \(q\) component of a vector field. The value measured is
\begin{align}
F^{q}_{\theta}\qty(s, t) = \int_{-\infty}^{\infty} &\hat{q}\cdot\vec{g}\qty( x\qty(q,s), y\qty(q,s) , t)  \dd{q} \\
= \int_{-\infty}^{\infty} [&g_y\qty( x\qty(q,s), y\qty(q,s) , t) \cos\theta \nonumber \\
 + &g_x\qty( x\qty(q,s), y\qty(q,s) , t) \sin\theta] \dd{q} \nonumber
\end{align}
It has been shown that these techniques may only be used to recover the (two-dimensionally) solenoidal part of a vector field~\cite{norton}, i.e. that which satisfies

\begin{equation}
\dv{g_x}{x} + \dv{g_y}{y} = 0
\end{equation}

Proton radiography represents a measurement of magnetic field components transverse to the probe direction, so may be used to implement a \emph{transverse vector tomography}. By this scheme the \(s\) and \(t\) components of a vector field are measured. These are given by
\begin{align}
F^{s}_{\theta}\qty(s, t) = \int_{-\infty}^{\infty} &\hat{s}\cdot\vec{g}\qty( x\qty(q,s), y\qty(q,s) , t) \dd{q} \\
= \int_{-\infty}^{\infty} &[g_x\qty( x\qty(q,s), y\qty(q,s) , t) \cos\theta \nonumber \\
+ &g_y\qty( x\qty(q,s), y\qty(q,s), t) \sin\theta] \dd{q} \nonumber
\end{align}
\begin{align}
F^{t}_{\theta}\qty(s, t) = \int_{-\infty}^{\infty} &\hat{t}\cdot\vec{g}\qty( x\qty(q,s), y\qty(q,s) , t) \dd{q} \\
= \int_{-\infty}^{\infty} &g_z\qty( x\qty(q,s), y\qty(q,s) , t) \dd{q} \nonumber
\end{align}
Only the irrotational part of the in-plane field is recoverable, which can be seen by following similar reasoning to the recoverability of the solenoidal part of the \(q\) component. This is unhelpful, when considering magnetic fields---the part of the field sourced by out-of-plane currents is undetectable, which can be seen from the relevant component of Amp\`ere's Law:

\begin{equation}
\mu_0 J_z = \pdv{B_y}{x} - \pdv{B_x}{y} = 0
\end{equation}

The component \(g_z\), in contrast, transforms as a scalar under the rotation which defines the probe geometry and is therefore accessible in full to traditional, scalar tomography techniques.

The following analysis will therefore focus on this out of plane component of the magnetic field---by rotating around a chosen axis, the component parallel to that axis is recoverable in full. Rotations about three orthogonal axes are therefore sufficient to recover the full three-dimensional vector field. This protocol reduces the problem of three-dimensional transverse vector tomography to a series of two-dimensional scalar tomography problems, allowing the use of well-developed algorithms and numerical techniques from this field.

\section{Derivation of the Angular Mode Approach}
\label{sec:hankel}

\subsection{The Fourier Projection-Slice Theorem}
\label{sec:hankel:fourier}

The Fourier Projection-Slice theorem states that the one-dimensional Fourier transform \(\tilde{F}_{\theta}\qty(k_s)\) of the projection \(F_{\theta}\qty(s)\) is equal to the two-dimensional Fourier transform \(\tilde{f}\qty(k_x, k_y)\) of the original function \(f\qty(x, y)\), evaluated on a one-dimensional slice through the origin of frequency space normal to the probe direction:

\begin{equation}
\tilde{F}_{\theta}\qty(k) = \tilde{f}\qty(-k \sin{\theta}, k \cos{\theta})
\end{equation}

Promoting the parameter \(\theta\) to a variable, it is clear that \(\tilde{F}\qty(k, \theta)\) is in fact nothing more than a representation of \(\tilde{f}\) in plane polar coordinates, albeit with \(\theta\) differing from the conventional polar angle variable by a quarter-cycle.

\begin{equation}
\tilde{F}\qty(k, \theta - \pi/2) = \tilde{f}\qty(k \cos{\theta}, k \sin{\theta})
\end{equation}

All that is in principle required for reconstruction of tomographic data is therefore a two-dimensional inverse Fourier transform of the one dimensional Fourier transform of projected data. This procedure is analytically exact, but complications arise due to the discrete sampling of real data. How should the inverse Fourier transform of data sampled discretely on a polar grid be computed?

Firstly, it is possible in general to represent a function discretely sampled in polar coordinates as a Fourier series in the angular variable (we shall now use \(\theta\) to denote angles in real space and \(\phi\) for those in the Fourier domain, to avoid confusion further into the derivation):

\begin{align}
\tilde{f}\qty(k_x, k_y) &= \tilde{F}\qty(k, \phi - \pi/2) \label{eq:Fseries}\\
&= \sum_{m=-M}^{M} \tilde{F}_m\qty(k) \exp(-i m (\phi  - \pi/2)), \nonumber
\end{align}
for a function sampled at M equally-spaced angles. 

It is now necessary to evaluate the inverse Fourier transform integral in two-dimensional polar coordinates. The integral is:

\begin{align}
\mathcal{F}^{-1}\left[\tilde{f}\right]\qty(r, \theta) &= \frac{1}{2\pi} \iint \exp(i \vec{k} \cdot \vec{r}) \tilde{f}\qty(\vec{k}) \dd^2{\vec{k}} \\
&= \frac{1}{2\pi}\int\limits_{0}^{\infty} \int\limits_{0}^{2\pi} [\exp(i kr\cos(\phi - \theta))\nonumber\\
&\qquad\qquad\qquad\tilde{F}\qty(k, \phi - \pi/2) k ] \dd{k}\dd{\phi}. \nonumber
\end{align}

Now, \(f\qty(r, \theta) = \mathcal{F}^{-1}[\tilde{f}]\) is also expanded as a Fourier series, analogously to Equation~\ref{eq:Fseries}:

\begin{equation}
f\qty(r, \theta) = \sum_{m=-M}^{M} f_m\qty(r) \exp(-i m \theta).
\end{equation}

Expanding both \(f\) and \(\tilde{F}\) into Fourier modes, defining \(\psi = \phi - \pi/2 - \theta\) and matching like terms in \(\exp(-im\theta)\) we obtain

\begin{align}
f_m\qty(r) &= \\
 &\frac{1}{2\pi}\int\limits_{0}^{\infty} \int\limits_{0}^{2\pi} \exp(-i\qty(kr\sin(\psi) + m\psi)) \dd{\phi} \tilde{F}_m\qty(k) k \dd{k} \nonumber
\end{align}

By comparison with a standard integral definition of the Bessel function of the first kind,

\begin{align}
J_m(x) &= \frac{1}{2\pi} \int\limits_{-\pi}^{\pi} \exp(i ( x\sin(\tau) - m \tau) ) \dd{\tau} \\
&= \frac{1}{2\pi} \int\limits_{0}^{2\pi} (-1)^m \exp(i ( -x\sin(\psi) - m \psi) ) \dd{\psi} \nonumber
\end{align}

allows the simplification:

\begin{align} \label{eq:Hankel}
f_m\qty(r) &= (-1)^m \int_{0}^{\infty} J_m\qty(kr) \tilde{F}_m\qty(k) k \dd{k} \\
&= (-1)^m \mathcal{H}_m\left[\tilde{F}_m\qty(k)\right]. \nonumber
\end{align}

The individual azimuthal Fourier modes of a two-dimensional function and its Fourier transform are related by a Hankel transform whose order is given by the Fourier mode number. This is a generalisation of the ``FHA cycle'', which states that the Fourier transform of the Abel transform of a one-dimensional function is equivalent to the function's Hankel transform of order 0. As the Abel transform corresponds to a Radon transform of a circularly symmetric function, the FHA cycle is the \(m = 0\) case of the above relation.

This suggests a procedure for recovery of functions from their projections \(F\left(s, \theta\right)\). First, a Fourier transform is performed along the \(s-\)axis and the result is decomposed into a Fourier series in \(\theta\). Then, a Hankel transform of the appropriate order is applied to each angular mode, yielding Fourier series components of the original function. This Fourier series is resolved with arbitrary angular resolution (as the angular dependence is given in terms of known functions \(\cos(\theta)\) and \(\sin(\theta)\)), allowing the reconstruction of the original function to be displayed in a smooth, visually appealing manner even when only a few sampling points are used.

\subsection{Discretisation and The Hankel Transform}

The procedure of the previous section involves the computation and resolution of Fourier series as well as the computation of Fourier and Hankel transforms. For discretely sampled data, the computation of both Fourier series and Fourier transforms are often efficiently carried out using the well-known Fast Fourier Transform (\texttt{FFT}) algorithm. The algorithm is therefore only missing one detail---computation of the Hankel Transform---before it may be implemented numerically. Numerous algorithms have been proposed to carry out this computation, including direct numerical quadrature; conversion of the Hankel transform to a convolution using a logarithmic change of variables; and methods using series expansions of the Bessel functions or of the function to be transformed~\cite{hankel,qfht}.

Methods relying on series expansion of the transformed function, for example into a sum of Bessel functions, require unevenly-sampled data so are inapplicable to the case at hand. Others involve adaptive Gauss quadrature, which requires knowledge of the analytic form of the function to be transformed. For discretely sampled data these methods are therefore also inapplicable. The remaining algorithms are direct (e.g. trapezium rule) integration of the Hankel transform integral Eq.~\ref{eq:Hankel}, and projection/back-projection methods.

The method for calculating integer-order Hankel transforms proposed by \textcite{hankel2} can be seen to be closely-related to the standard filtered back-projection (FBP) algorithm, in the limit where the discrete sum in FBP becomes a continuous angular integral (high sampling rate) and the angular dependence of the integrand/summand is given by \(\exp(i m \theta)\) (an angular Fourier mode). Therefore, the method of Higgins and Munson is approximated for any of the modes in Eq.~\ref{eq:Hankel} by passing a filtered back-projection algorithm a set of `virtual projections' with the desired angular dependence, sampled with arbitrary angular density. The sum over all modes of these virtual projections matches the true projections exactly at the original sampling points and their values will be \(C^\infty\) smooth between sample points. The filtered back-projection of the virtual projections of each mode then returns the corresponding Fourier series component of the original, real-space function \(f_m\), including the \(\exp(i m \theta)\) angular dependence.

Resolving the Fourier series of the real-space function is therefore simple:
\begin{equation}
	f\qty(x, y) \approx \sum\limits_{m=-M}^{M} \operatorname{FBP}\qty[ F_m\qty(k) \exp(i m \phi) ]. \label{eq:FBP}
\end{equation}

By linearity of the Radon transform (which is inherited by filtered back-projection) the summation in Equation~\ref{eq:FBP} may be moved inside of the FBP computation:

\begin{align}
f\qty(x, y) &\approx \operatorname{FBP}\qty[ \sum\limits_{m=-M}^{M} F_m\qty(k) \exp(i m \phi) ] \\
    &= \operatorname{FBP}\qty[ \hat{F}\qty(k, \phi) ] \nonumber
\end{align}

This equation provides a useful interpretation of \(\hat{F}\): it is a version of \(F\) `enhanced' by interpolation in \(\phi\) using its Fourier series. As noted above, the sum over modes \(m\) of the Fourier series agrees with the original samples \(F\) at the sampled angles and is `optimal' in the sense that it is \(C^\infty\)-smooth and explicitly possesses the same \(2\pi\)-periodicity that the true function must exhibit.

The observation that the Fourier series is resolved immediately by `pushing' the final summation inside the filtered back-projection has several advantages.

Firstly it allows a simplification of the algorithm presented in the previous section, by facilitating use of existing implementations of filtered back-projection. As FBP acts directly on tomographic observations rather than their Fourier transforms the \texttt{FFT} calculation along \(s\) is no longer needed. This \texttt{FFT} may happen inside the FBP algorithm to achieve its filtering of samples, but it could also be implemented using a direct convolution. The FBP algorithm is free to choose dynamically how to perform it, using speed heuristics for example as seen in the \texttt{scipy} function \texttt{scipy.signal.convolve}~\cite{scipy}.

Secondly, the interpretation of the resulting algorithm as the application of a standard tomographic reconstruction algorithm to data enhanced using trigonometric interpolation implies that algorithms other than FBP could be used to implement that stage of the process. 

\subsection{Demonstration of Fourier Interpolated Tomographic Reconstruction}

\begin{figure}
	\includegraphics[width=\columnwidth]{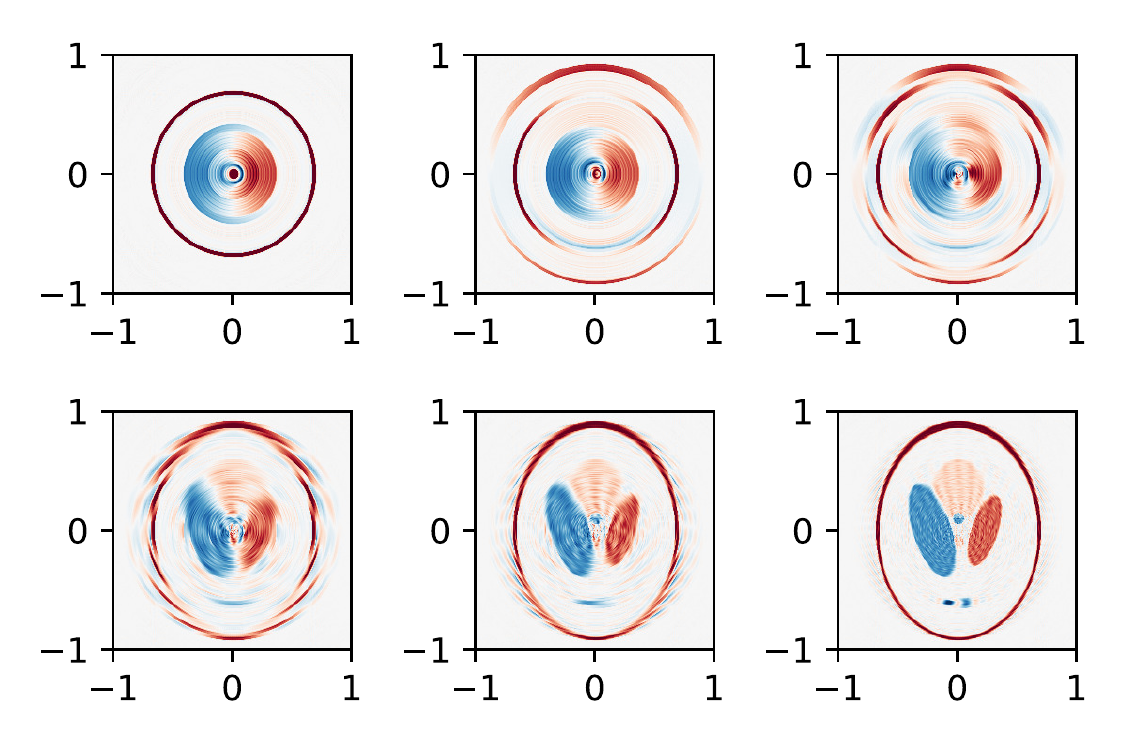}
	\caption{Demonstration of the improvement attainable by using Fourier interpolation techniques. Images were produced identically to Figure~\ref{fig:fbp_streaks}, except that the set of projections was enhanced by interpolation to an angular frequency of 1024 view angles prior to reconstruction. Noise and streaking are significantly reduced, though performance is still quite poor for very small numbers of projections. Linear streaking has been replaced by circular streaking, reflecting this method's polar-coordinates formulation as opposed to the Cartesian formulation of standard FBP.}
	\label{fig:fourier_streaks}
\end{figure}

Figure~\ref{fig:fbp_streaks} was recreated using the Fourier interpolation technique to pre-enhance the set of all projections. The result is shown in Figure~\ref{fig:fourier_streaks}. It can be seen that in all cases the properties of the interpolated reconstruction are improved versus the non-interpolated. For example, while the na\"ive application of filtered backprojection in Figure~\ref{fig:fbp_streaks} results in functions featuring streaks that reach the edges of the image (and would continue arbitrarily far if the reconstruction was carried out on a larger domain), the reconstruction of Fourier-interpolated images is compactly supported on the smallest disk that completely contains the support of the original function. Effectively, the linear streaks seen in the FBP reconstructions correspond to circular streaks of constant radius in the reconstructed image, and these circular streaks will--for many functions--represent less of a deviation from the original function than the linear streaks characteristic of FBP in the sparse sampling regime. This correspondence reflects the relationship between standard FBP and the polar-coordinates approach taken in the derivation of this method: artefacts in FBP arise from `smearing' the observations along the direction of observation; artefacts in Fourier-interpolated FBP represent a `smearing' in the polar angle.

Further, the most noticeable artefacts in, for example, the reconstruction of Figure~\ref{fig:fourier_streaks}, panel c) appear in the region where the elliptical support of the original function does not fill the circular support of the reconstruction. It is therefore likely that the reconstruction would be improved further by employing a virtual transformation to the computational domain of the algorithm to ensure the best possible overlap between the original function's region of support and the disk containing that region. In the next section, this conjecture is investigated and a method is derived to achieve the necessary transformation.

\section{Aspect Ratio Compensation: Elliptical Tomography}
\label{sec:aspect}

\begin{figure}
	\includegraphics[width=\columnwidth]{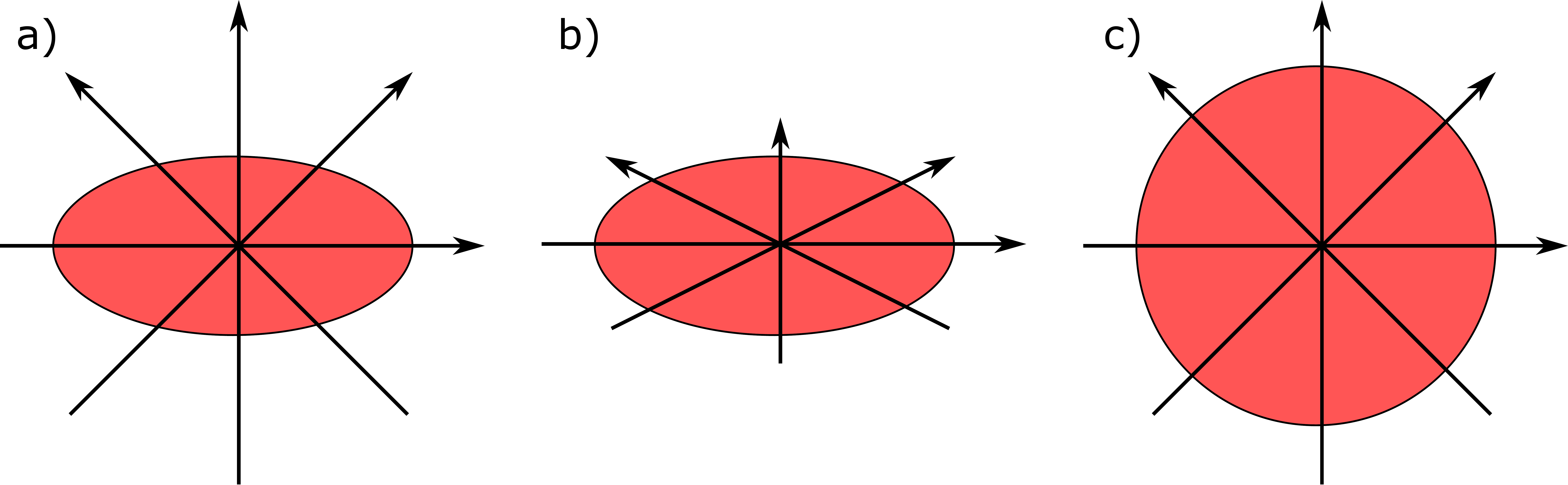}
	\caption{Schematic depiction of the aspect ratio compensation method for an elliptical target function of aspect ratio 2:1. a) uniform sampling of the elliptical target function by four probes. b) sampling at angles given by Equation~\ref{eq:angle_compensation} for a compensation ratio of 2:1, showing non-uniform sampling in the physical domain. c) transformation into the computational domain of subfigure b), showing that in the computational domain sampling is uniform in angle and the target has been transformed into a circle.}
	\label{fig:aspect_schematic}
\end{figure}

Many realistic functions are not best described as being supported on a disk, but have some aspect ratio not equal to unity. These functions, represented by their projections as a sinogram, have a width that oscillates with angle, an effect seen for example in the right panel of Figure~\ref{fig:sinogram}. Altering the procedure used such that the object appears to have aspect ratio closer to unity may be expected to improve the reconstruction quality. This may be achieved by applying a single-axis scaling between the physical and computational domains of the problem. There are three things which must be considered when implementing such a scaling: first, the angular separation of observations in physical space becomes non-uniform in order to maintain uniform angular sampling in computational space; second, an individual scaling of the \(s\)-axis must be applied to projections, accounting for the stretching or shrinking of the axis perpendicular to the projection; and third, this must be compensated for using an inverse scaling of the function values to maintain equality of all projections' integrals along \(s\).

The modified Shepp-Logan phantom used in Figures~\ref{fig:fbp_streaks} and \ref{fig:fourier_streaks} is defined on the two-dimensional domain \([-1,1]\times[-1,1]\) and composed of several ellipses of differing parameters. The support of this function is defined by the largest ellipse, which entirely contains all other ellipses and has minor and major semi-axis lengths in the ratio \(A = 3/4\). We now detail the procedure for tomography of this phantom using aspect ratio compensation between the physical domain with coordinates \((X, Y; R, \Theta)\) and the computational domain with coordinates \((x, y; r, \theta)\).

The relation between Cartesian coordinates is such as to equalise the aspect ratio of the function under observation. Keeping \(x = X\), this implies \(y = Y / A\). Angles of projection are uniformly spaced in the computational domain: 

\begin{equation}
\theta_k = \frac{k \pi}{N}\quad(0 \leq k < N)
\end{equation}

Using the relationship between Cartesian coordinates it is easy to derive the corresponding relationship for the angular variables \(\Theta\) and \(\theta\):

\begin{equation}
\tan\Theta = \frac{Y}{X} = \frac{Ay}{x} = A\tan{\theta} \label{eq:angle_compensation}
\end{equation}

This has the effect of reducing angular spacing when the probe direction is close to the major axis and increasing angular spacing when close to the minor axis.

The transverse extent of the physical-space object varies with viewing angle, and this must also be compensated for. The physical transverse width of the ellipse is 

\begin{align}
w(\Theta) &= \sqrt{w_\mathrm{max}^2 \cos^2{\Theta} + w_\mathrm{min}^2 \sin^2{\Theta}} \label{eq:wtheta}\\
&= w_\mathrm{max}\cos{\Theta} \sqrt{1 + A^{-2}\tan^2{\Theta}} \nonumber \\
&= w_\mathrm{max}\cos{\Theta} \sqrt{1 + \tan^2{\theta}} \nonumber \\
&= w_\mathrm{max}\frac{\cos{\Theta}}{\cos{\theta}}. \nonumber
\end{align}

\begin{figure*}
	\includegraphics[width=\textwidth]{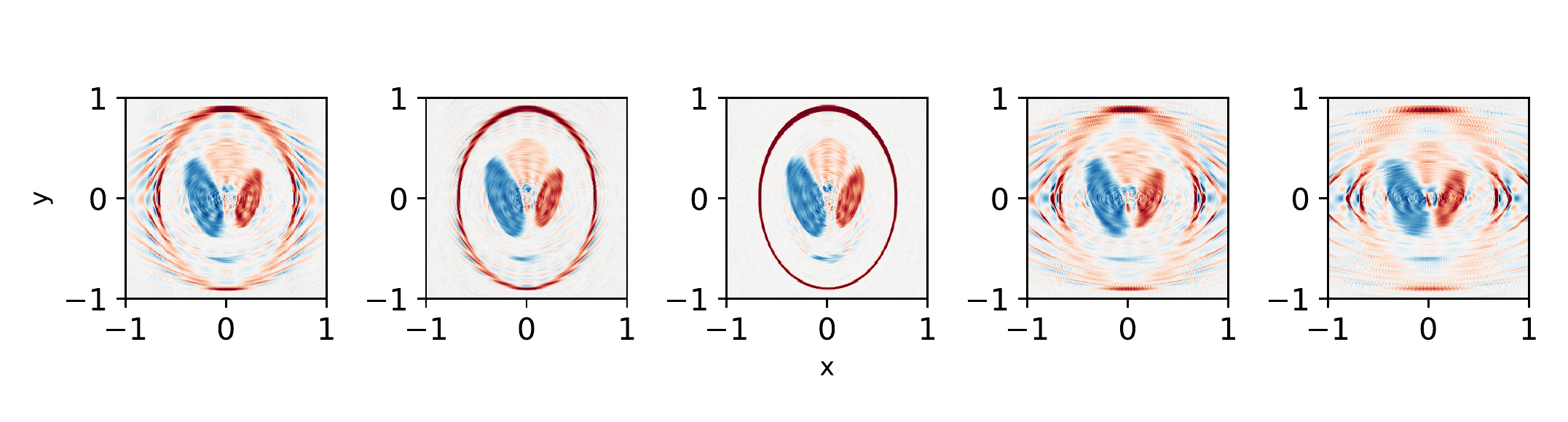}
	\caption{The effect of aspect ratio compensation on reconstruction quality. All reconstructions were carried out using 15 projections, and expected aspect ratios of a) 4:3, b) 1:1, c) 3:4, d) 9:16, e) 27:64 chosen as they are integer powers of the correct 3:4 ratio. b) shows no compensation; c) shows the correct level of compensation and resultantly the highest quality of reconstruction; a) shows the correct ratio but in the wrong direction, worsening the quality of the reconstruction; d) and e) `squeeze' the function due to excessive compensation. It is clear that compensation for aspect ratio can improve reconstruction quality, if applied appropriately.}
	\label{fig:aspect}
\end{figure*}

The \(s\)-axis of each projection is re-scaled by a factor \(w(\Theta) / w_\mathrm{min} = A \cos{\Theta} / \cos{\theta}\) to account for the transverse stretching caused by aspect ratio correction, and the magnitudes of each projection's values are scaled by the inverse value (Equation \ref{eq:wtheta} is used preferentially as the limit \(\Theta=\theta=\pi/2\) is not problematic in this form). This has the effect of eliminating the oscillation of the sinogram's width as a function of \(\Theta\). The techniques detailed above in Section~\ref{sec:tomography} may then be applied to this modified sinogram and the end result of the reconstruction is stretched to apply the physically correct aspect ratio. The result of this procedure is shown in Figure~\ref{fig:aspect}.

\section{Applications}
\label{sec:applications}

\begin{figure*}
	\includegraphics[width=0.5\textwidth]{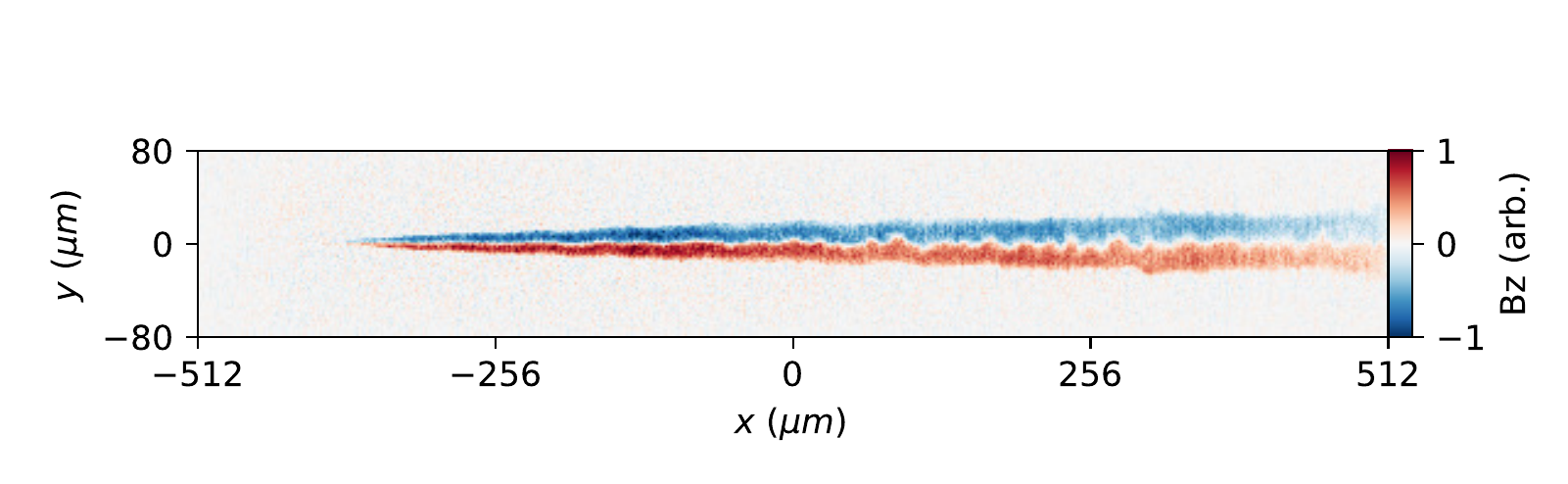}\\
	\includegraphics[width=\textwidth]{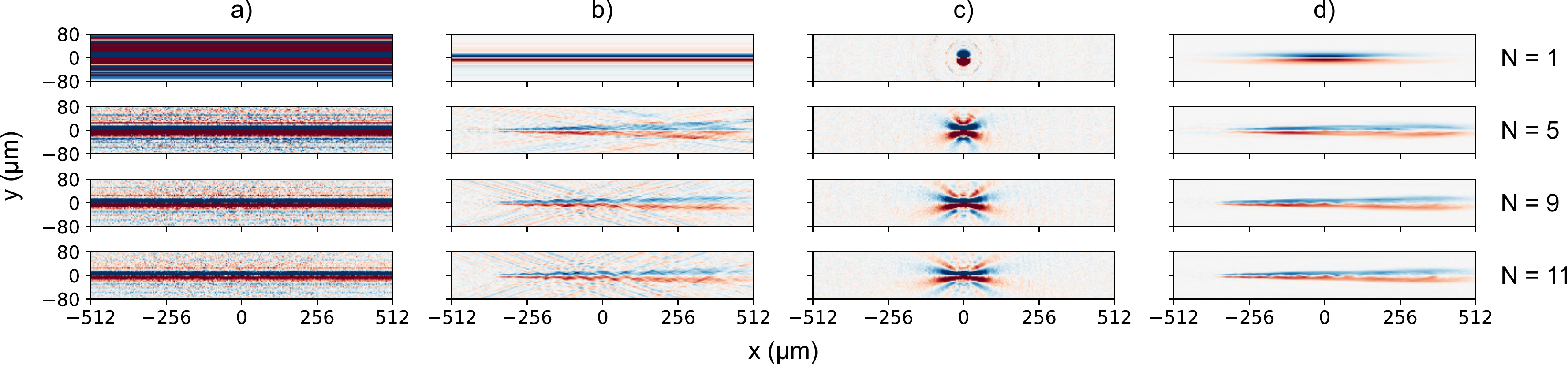}
	\caption{Above: Z-component of magnetic field extracted from a particle-in-cell simulation of a laser-plasma channelling interaction. The field is shown in two dimensions, averaged over the z-axis of the simulation to improve legibility. All of the methods presented here are however applicable to three-dimensional data sets as well as two-dimensional ones. Below: Demonstration of this paper's proposed techniques at a range of sampling rates. \(N\) denotes the number of angular observations included in the reconstruction. Column a) presents the results of applying na\"ive filtered back-projection, b) includes aspect ratio-compensation, c) includes Fourier-series interpolation and d) includes both. Aspect ratio compensation uses a ratio of 10:1. While each technique struggles with this field when used individually, the composition of both techniques produces results of good quality even for very sparse sampling. All reconstructions are plotted using the same colour scaling as the original field.}
	\label{fig:channel_comparison}
\end{figure*}

We now turn our attention to the important example of imaging laser-plasma interactions with very large aspect ratios, such as channelling processes~\cite{spiers}, jets in laboratory plasma astrophysics experiments~\cite{chikang3} and z-pinches~\cite{lebedev}. To demonstrate its utility for the first of these applications, magnetic fields have been extracted from a particle-in-cell simulation of a high-intensity laser pulse propagating into a plasma with a pre-formed density gradient. A representation of these fields is shown in figure \ref{fig:channel_comparison}. 

The results of reconstructing this field with and without both Fourier interpolation and 10:1 aspect ratio compensation are shown in Figure~\ref{fig:channel_comparison}, using a range of angular sampling rates to show the deterioration in reconstruction quality for each method as the number of observations is reduced. The highly elongated nature of the field displayed in Figure~\ref{fig:channel_comparison} causes severe problems in the absence of aspect ratio compensation, though Fourier-series interpolation improves the appearance of the final result. Even with aspect-ratio compensation applied, without Fourier interpolation the result still suffers from streaking artefacts which can obscure the true field. At all sampling rates tested, applying both techniques together performs better than either individually, demonstrating that the noise- and artefact-reduction properties of Fourier-series interpolation complement the more efficient sampling of Fourier space allowed by aspect ratio compensation.

\section{Summary and Conclusions}
\label{sec:conclusion}

Proton radiography has found many applications for probing magnetic field structures in plasma. However, its extension to three dimensional reconstruction remains a significant challenge. To this end, two novel pre-processing techniques for improving the performance of a standard tomographic reconstruction algorithm---filtered back-projection---have been explored in this article.

First, Fourier decomposition of observations in the angular parameter was proposed as a method for exact inversion of the Radon transform, based on the generalised Fourier-Hankel-Abel cycle of integral transforms which derives from the Fourier projection-slice theorem. By approximating the calculation of the general integer-order Hankel transform using back-projection, one observes that a single filtered back-projection of interpolated data is able to replace the calculation of a different integer-order Hankel transform per angular mode, greatly reducing the computational complexity of the method.

Second, based on the properties of Fourier-interpolated reconstructions, this method of tomography has been shown to achieve better accuracy for small numbers of observations when the aspect ratio of the function being observed is close to unity. To benefit from this observation, relations linking physical space and a computational space which differ by a non-uniform scaling have been derived, and these relations allow aspect ratios far from unity to be compensated for. The effectiveness of this compensation technique has been demonstrated using a modified Shepp-Logan phantom, which is supported on an ellipse of aspect ratio 3:4.

The effectiveness of these new proposed pre-processing enhancement steps, both individually and in combination has been compared to `pure' filtered back-projection. It has been shown that in the case of the magnetic field of a simulated laser channel in dense plasma, each new pre-processing method improves the quality of reconstruction, and that combining them produces the best results of all. This significantly improves the prospects of a tomographic approach to proton radiography being implemented.

Finally, one notes that the methods presented here are also applicable to other path-integrated plasma probe diagnostics~\cite{kasim1} that have applications across the natural sciences and engineering.

\paragraph{Acknowledgements}

This work has been carried out within the framework of the EUROfusion Consortium and has received funding from the Euratom research and training programme 2019-2020 under grant agreement No 633053, with support of STFC grant ST/P002048/1 and EPSRC grants EP/R029148/1 and EP/L000237/1. The views and opinions expressed herein do not necessarily reflect those of the European Commission. BTS acknowledges support from UKRI-EPSRC and AWE plc. PAN acknowledges support from OxCHEDS for his William Penney Fellowship. The simulations presented herein were carried out using the ARCHER2 UK National Supercomputing Service. The authors gratefully acknowledge the support of all the staff at the Central Laser Facility, UKRI-STFC Rutherford Appleton Laboratory and the ORION Laser Facilty at AWE Aldermaston (particularly discussions with Gavin Crow) while undertaking this research.

\bibliography{predraft}
\end{document}